\documentclass[conference,a4paper]{IEEEtran}

\renewcommand\IEEEkeywordsname{Index Terms}

\usepackage{amsmath,amssymb,bm}
\usepackage{graphicx}
\usepackage{diagbox}
\graphicspath{{fig/}}
\usepackage{booktabs}
\usepackage{multirow}
\usepackage{tabularx, nicematrix}
\usepackage[caption=false,font=footnotesize]{subfig}
\usepackage{cite}
\usepackage[hyphens]{url}
\usepackage[bookmarks=false]{hyperref}
\usepackage[nolist]{acronym}
\usepackage{xcolor}

\usepackage{tikz}
\usetikzlibrary{arrows.meta,positioning,fit,backgrounds,decorations.pathreplacing}
\usetikzlibrary{decorations.text}

\usepackage{pgfplots}
\pgfplotsset{compat=1.18}

\pgfdeclareplotmark{plus*}{%
  \pgfpathmoveto{\pgfqpoint{-\pgfplotmarksize}{-.33\pgfplotmarksize}}%
  \pgfpathlineto{\pgfqpoint{-\pgfplotmarksize}{.33\pgfplotmarksize}}%
  \pgfpathlineto{\pgfqpoint{-.33\pgfplotmarksize}{.33\pgfplotmarksize}}%
  \pgfpathlineto{\pgfqpoint{-.33\pgfplotmarksize}{\pgfplotmarksize}}%
  \pgfpathlineto{\pgfqpoint{.33\pgfplotmarksize}{\pgfplotmarksize}}%
  \pgfpathlineto{\pgfqpoint{.33\pgfplotmarksize}{.33\pgfplotmarksize}}%
  \pgfpathlineto{\pgfqpoint{\pgfplotmarksize}{.33\pgfplotmarksize}}%
  \pgfpathlineto{\pgfqpoint{\pgfplotmarksize}{-.33\pgfplotmarksize}}%
  \pgfpathlineto{\pgfqpoint{.33\pgfplotmarksize}{-.33\pgfplotmarksize}}%
  \pgfpathlineto{\pgfqpoint{.33\pgfplotmarksize}{-\pgfplotmarksize}}%
  \pgfpathlineto{\pgfqpoint{-.33\pgfplotmarksize}{-\pgfplotmarksize}}%
  \pgfpathlineto{\pgfqpoint{-.33\pgfplotmarksize}{-.33\pgfplotmarksize}}%
  \pgfpathclose
  \pgfusepathqfillstroke
}

\usepgfplotslibrary{groupplots}
\usepgfplotslibrary{fillbetween}

\usetikzlibrary{spy,calc}

\definecolor{mittelblau}{RGB}{0, 126, 198}
\definecolor{violettblau}{cmyk}{0.9, 0.6, 0, 0}
\definecolor{rot}{RGB}{238, 28 35}
\definecolor{apfelgruen}{RGB}{140, 198, 62}
\definecolor{gelb}{RGB}{255, 229, 0}
\definecolor{orange}{RGB}{244, 111, 33}
\definecolor{pink}{RGB}{237, 0, 140}
\definecolor{lila}{RGB}{128, 10, 145}
\definecolor{hellgrau}{RGB}{224, 224, 224}
\definecolor{mittelgrau}{RGB}{128, 128, 128}
\definecolor{dunkelgrau}{RGB}{80,80,80}
\definecolor{anthrazit}{RGB}{19, 31, 31}
\definecolor{darkgreen}{RGB}{34,139,34}
\definecolor{aqua}{RGB}{0, 255, 255}

\definecolor{lightgray}{RGB}{211,211,211}

\definecolor{neuesgruen}{RGB}{61, 173, 65}
\definecolor{dunklereshellgrau}{RGB}{176, 176, 176}
\definecolor{neuesgelb}{RGB}{255,160,0}
\definecolor{neuescyan}{RGB}{69,185,224}

\definecolor{tollesgruen}{RGB}{0,217,171}
\definecolor{tollesmagenta}{RGB}{197,67,143}

\definecolor{tollesgelb}{RGB}{255,199,95}
\definecolor{tollesrot}{RGB}{255,111,145}

\colorlet{R12}{apfelgruen}
\colorlet{R23}{mittelblau}
\colorlet{R45}{pink}

\renewcommand{\vec}[1]{\boldsymbol{#1}}

\newcommand{\Lm}{\vec{L}}

\tikzset{
       vnd/.style={
        shape=circle,
        fill=black,
        draw,
        inner sep=0pt,
        minimum size=0.2cm},
        cnd/.style={
        shape=rectangle,
        fill=white,
        draw,
        minimum width=0.05mm,
        minimum height = 0.05mm}, 
         vndR/.style={
        shape=circle,
        fill=red,
        draw,
        inner sep=0pt,
        minimum size=0.2cm},
        cndR/.style={
        shape=rectangle,
        fill=white,
        draw=red,
        minimum width=0.05mm,
        minimum height = 0.05mm}
}

\newcommand{\comb}{\mathrm{comb}}
\newcommand{\Lcomb}{\Lm_{\comb}}               %
\newcommand{\Lch}{\Lm_\mathrm{ch}}   %
\DeclareMathOperator{\clip}{clip}

\IEEEoverridecommandlockouts
\begin{document}
\setlength{\columnsep}{0.21in}

\title{SECS: A Soft Ensemble-Combining Stage for Low-Latency Decoding}

\author{\IEEEauthorblockN{Felix Krieg\textsuperscript{\dag}, Paul Bezner\textsuperscript{\dag}, and Stephan ten Brink}
\IEEEauthorblockA{Institute of Telecommunications, University of Stuttgart, Germany\\
\{krieg, bezner, tenbrink\}@inue.uni-stuttgart.de}
\thanks{\textsuperscript{\dag}\,F. Krieg and P. Bezner contributed equally to this work.}
\thanks{This work is supported by the German Federal Ministry of Research, Technology and Space (BMFTR) within the project Open6GHub+ (grant no. 16KIS2406).}
\thanks{LLM agents from Anthropic and OpenAI were used to write source code for this work and to assist with language editing.}}

\maketitle

\begin{acronym}
\acro{SECS}{soft ensemble-combining stage}
\acro{AED}{automorphism ensemble decoding}
\acro{AVN}{auxiliary variable node}
\acro{AWGN}{additive white Gaussian noise}
\acro{BCH}{Bose--Ray-Chaudhuri--Hocquenghem}
\acro{BDD}{bounded-distance decoding}
\acro{BER}{bit error rate}
\acro{BKLC}{best known linear code}
\acro{BP}{belief propagation}
\acro{BPSK}{binary phase-shift keying}
\acro{CRC}{cyclic redundancy check}
\acro{FER}{frame error rate}
\acro{GE}{Gaussian elimination}
\acro{HDPC}{high-density parity-check}
\acro{LDPC}{low-density parity-check}
\acro{LER}{list error rate}
\acro{LLR}{log-likelihood ratio}
\acro{LRB}{least-reliable basis}
\acro{MBBP}{multiple-bases belief propagation}
\acro{ML}{maximum likelihood}
\acro{MRB}{most-reliable basis}
\acro{OSD}{ordered-statistics decoding}
\acro{PCM}{parity-check matrix}
\acro{QR}{quadratic-residue}
\acroplural{PCM}[PCMs]{parity-check matrices}
\acro{RM}{Reed--Muller}
\acro{SNR}{signal-to-noise ratio}
\acro{CRT}{Chinese remainder theorem}
\acro{SC}{successive cancellation}
\acro{ssPCM}{structured sparse parity-check matrix}
\acroplural{ssPCM}[ssPCMs]{structured sparse parity-check matrices}
\acro{URLLC}{ultra-reliable low-latency communication}
\acro{NMSA}{normalized min--sum algorithm}
\acro{SPA}{sum--product algorithm}
\acro{XOR}{exclusive OR}
\end{acronym}

\begin{abstract}
Ensemble decoding is a promising technique for \acl{URLLC}, as it trades
hardware parallelism for decoding latency by running \(M\) diverse
\ac{BP} decoders in parallel.
Conventional ensembles select their output from the member candidates.
Hence, decoding fails whenever no member finds the correct codeword.
In this work, we show that shallow, diverse \ac{BP} members are poor
\emph{correctors}, but excellent \emph{sorters}.
Based on this observation, we propose a \ac{SECS} that
combines extrinsic messages after only a few iterations, yielding a
reliability ordering whose most reliable positions are nearly error-free.
A subsequent re-encoding stage (like \ac{OSD}) converts this ordering into a near-\ac{ML}
candidate codeword, at a fraction of the \ac{BP}
latency of a fully converged ensemble.
We demonstrate the proposed scheme on three short codes: a
\((63,30)\) \acl{BCH} code, an overcomplete \(\mathrm{PG}(2,8)\) code, and
a search-designed \((105,53)\) cyclic code.
In all cases, the \ac{SECS} with \ac{OSD} post-processing
closes most of the gap to \ac{ML} decoding, while significantly reducing
the number of required \ac{BP} iterations.
\end{abstract}

\acresetall

\section{Introduction}
\label{sec:intro}

\begin{figure}[!t]
\centering
\begin{tikzpicture}
\begin{groupplot}[
    group style={group size=2 by 1, horizontal sep=7mm,
                 ylabels at=edge left, yticklabels at=edge left},
    width=0.49\linewidth, height=0.5\linewidth,
    ymode=log, log origin=infty, grid=both,
    major grid style={draw=hellgrau!85, line width=0.36pt},
    minor grid style={draw=hellgrau!45, line width=0.22pt},
    axis x line*=bottom, axis y line*=left,
    axis line style={draw=anthrazit, line width=0.50pt},
    tick align=outside, tick style={draw=anthrazit, line width=0.45pt},
    ticklabel style={font=\footnotesize, text=anthrazit},
    label style={font=\footnotesize, text=anthrazit},
    title style={font=\footnotesize, text=anthrazit, yshift=-1mm},
    xmin=0, xmax=62, ymin=1e-06, ymax=1,
    xlabel={bit reliability rank \(r\)},
    ylabel={BER at rank \(r\)},
]

\nextgroupplot[title={\(3.0\,\)dB}]
\addplot+[color=dunkelgrau, solid, line width=0.9pt, mark=none] coordinates {(0,6.395e-05) (1,0.0001544) (2,0.0002473) (3,0.0003656) (4,0.0004785) (5,0.0006496) (6,0.0008274) (7,0.00103) (8,0.001245) (9,0.001497) (10,0.001777) (11,0.002096) (12,0.002442) (13,0.002852) (14,0.003288) (15,0.003762) (16,0.004277) (17,0.004841) (18,0.005519) (19,0.006248) (20,0.007055) (21,0.007885) (22,0.008859) (23,0.009929) (24,0.01106) (25,0.01236) (26,0.01373) (27,0.01529) (28,0.01708) (29,0.01893) (30,0.02098) (31,0.0232) (32,0.02579) (33,0.02857) (34,0.03163) (35,0.03503) (36,0.03879) (37,0.04296) (38,0.04754) (39,0.05259) (40,0.0581) (41,0.06435) (42,0.07121) (43,0.0788) (44,0.08704) (45,0.09641) (46,0.1068) (47,0.1179) (48,0.1302) (49,0.1442) (50,0.1591) (51,0.1754) (52,0.1935) (53,0.2133) (54,0.2346) (55,0.2575) (56,0.2825) (57,0.3092) (58,0.3375) (59,0.3676) (60,0.3989) (61,0.4316) (62,0.4656)};
\addplot+[color=mittelblau, solid, line width=0.9pt, mark=none] coordinates {(0,1.45e-05) (1,3e-05) (2,5.6e-05) (3,8e-05) (4,0.000117) (5,0.0001405) (6,0.000188) (7,0.00025) (8,0.000293) (9,0.000357) (10,0.0004665) (11,0.000524) (12,0.0006325) (13,0.000742) (14,0.0008835) (15,0.001014) (16,0.001135) (17,0.001328) (18,0.001559) (19,0.001745) (20,0.001941) (21,0.002224) (22,0.00252) (23,0.002962) (24,0.003223) (25,0.003631) (26,0.00411) (27,0.004621) (28,0.005182) (29,0.005855) (30,0.006537) (31,0.007276) (32,0.008185) (33,0.009017) (34,0.01012) (35,0.01126) (36,0.01265) (37,0.01416) (38,0.01577) (39,0.01749) (40,0.01952) (41,0.02201) (42,0.02434) (43,0.02711) (44,0.03033) (45,0.03373) (46,0.03751) (47,0.04186) (48,0.04678) (49,0.05252) (50,0.05857) (51,0.06568) (52,0.0734) (53,0.08245) (54,0.09288) (55,0.1042) (56,0.1177) (57,0.1325) (58,0.1511) (59,0.1725) (60,0.1992) (61,0.2346) (62,0.2877)};
\addplot+[color=apfelgruen, solid, line width=0.95pt, mark=none] coordinates {(0,6.8e-05) (1,0.0001025) (2,0.000114) (3,0.000137) (4,0.0001465) (5,0.0001985) (6,0.0002145) (7,0.000247) (8,0.000283) (9,0.000332) (10,0.000362) (11,0.0004405) (12,0.000498) (13,0.0005685) (14,0.0006265) (15,0.000694) (16,0.0007765) (17,0.0008595) (18,0.000957) (19,0.001037) (20,0.001131) (21,0.001279) (22,0.001429) (23,0.001572) (24,0.001721) (25,0.001924) (26,0.00207) (27,0.002311) (28,0.002531) (29,0.002772) (30,0.002962) (31,0.003295) (32,0.003569) (33,0.003855) (34,0.004082) (35,0.004483) (36,0.004946) (37,0.005319) (38,0.005772) (39,0.006124) (40,0.006518) (41,0.007059) (42,0.007527) (43,0.008023) (44,0.008486) (45,0.008867) (46,0.009426) (47,0.009915) (48,0.01044) (49,0.01083) (50,0.01152) (51,0.01189) (52,0.01238) (53,0.01295) (54,0.01351) (55,0.01385) (56,0.01437) (57,0.01503) (58,0.0154) (59,0.01592) (60,0.01638) (61,0.01703) (62,0.01732)};
\addplot+[color=rot, solid, line width=1.4pt, mark=none] coordinates {(0,3.2e-06) (1,7.75e-06) (2,1.48e-05) (3,2.295e-05) (4,2.835e-05) (5,4.08e-05) (6,5.555e-05) (7,7.145e-05) (8,8.555e-05) (9,0.0001034) (10,0.0001323) (11,0.000159) (12,0.0001924) (13,0.0002185) (14,0.0002638) (15,0.0003008) (16,0.0003425) (17,0.0004009) (18,0.0004743) (19,0.0005307) (20,0.0006063) (21,0.0006863) (22,0.0007818) (23,0.0008935) (24,0.001017) (25,0.001152) (26,0.001313) (27,0.001482) (28,0.001646) (29,0.001869) (30,0.002084) (31,0.002358) (32,0.002657) (33,0.002952) (34,0.003314) (35,0.00369) (36,0.004147) (37,0.004656) (38,0.005173) (39,0.005791) (40,0.006448) (41,0.007233) (42,0.008108) (43,0.009042) (44,0.01007) (45,0.01128) (46,0.0126) (47,0.01409) (48,0.01577) (49,0.01762) (50,0.01977) (51,0.02219) (52,0.02491) (53,0.0279) (54,0.03142) (55,0.03536) (56,0.03997) (57,0.04515) (58,0.05134) (59,0.05909) (60,0.06889) (61,0.08285) (62,0.1086)};
\draw[anthrazit, densely dashed, line width=0.7pt] (axis cs:30,1e-07) -- (axis cs:30,1);
\node[anthrazit, font=\scriptsize, anchor=west] at (axis cs:30,5.0e-06) {\(k{=}30\)};
\nextgroupplot[title={\(3.5\,\)dB}, legend to name=ranklegendx,
      legend columns=4,
      legend cell align={left},
      legend image code/.code={
          \draw[mark repeat=2, mark phase=2, #1]
              plot coordinates {(0cm,0cm) (0.18cm,0cm) (0.36cm,0cm)};
      },
      legend style={
          draw=black,
          fill=white,
          font=\scriptsize,
          inner xsep=2pt,
          inner ysep=2pt,
          /tikz/every even column/.append style={column sep=0.18cm}
      },]
\addplot+[color=dunkelgrau, solid, line width=0.9pt, mark=none] coordinates {(0,3.07e-05) (1,7.105e-05) (2,0.0001195) (3,0.0001783) (4,0.0002525) (5,0.0003343) (6,0.0004277) (7,0.000538) (8,0.0006681) (9,0.0008167) (10,0.0009645) (11,0.001156) (12,0.001361) (13,0.001591) (14,0.001863) (15,0.002149) (16,0.00249) (17,0.002824) (18,0.003228) (19,0.003689) (20,0.004187) (21,0.004741) (22,0.005339) (23,0.006009) (24,0.00678) (25,0.007593) (26,0.008547) (27,0.009604) (28,0.01074) (29,0.01202) (30,0.01349) (31,0.01507) (32,0.0169) (33,0.01882) (34,0.021) (35,0.02349) (36,0.02625) (37,0.02929) (38,0.0328) (39,0.03668) (40,0.041) (41,0.0459) (42,0.05133) (43,0.05757) (44,0.06438) (45,0.07201) (46,0.08075) (47,0.09064) (48,0.1017) (49,0.114) (50,0.1279) (51,0.1435) (52,0.1608) (53,0.1802) (54,0.2018) (55,0.2256) (56,0.2519) (57,0.2808) (58,0.312) (59,0.3456) (60,0.3816) (61,0.4194) (62,0.459)};
\addplot+[color=mittelblau, solid, line width=0.9pt, mark=none] coordinates {(0,3.5e-06) (1,8.5e-06) (2,1.25e-05) (3,1.9e-05) (4,3.5e-05) (5,4.75e-05) (6,5.1e-05) (7,5.8e-05) (8,8.9e-05) (9,0.0001055) (10,0.00014) (11,0.0001585) (12,0.0001905) (13,0.000226) (14,0.0002665) (15,0.000332) (16,0.000362) (17,0.000424) (18,0.000484) (19,0.000567) (20,0.0006585) (21,0.000775) (22,0.0009295) (23,0.000995) (24,0.001133) (25,0.001252) (26,0.001477) (27,0.001689) (28,0.001901) (29,0.002139) (30,0.002457) (31,0.002738) (32,0.003128) (33,0.003498) (34,0.0039) (35,0.004497) (36,0.005128) (37,0.005685) (38,0.006464) (39,0.007373) (40,0.008372) (41,0.009331) (42,0.01065) (43,0.0119) (44,0.01361) (45,0.01542) (46,0.01753) (47,0.01974) (48,0.02236) (49,0.02551) (50,0.02894) (51,0.03333) (52,0.03779) (53,0.04328) (54,0.04955) (55,0.05712) (56,0.06622) (57,0.07599) (58,0.08908) (59,0.1048) (60,0.1252) (61,0.1538) (62,0.2004)};
\addplot+[color=apfelgruen, solid, line width=0.95pt, mark=none] coordinates {(0,2.3e-05) (1,3.65e-05) (2,4.35e-05) (3,5.2e-05) (4,5e-05) (5,6.85e-05) (6,7.45e-05) (7,8.9e-05) (8,0.000108) (9,0.000115) (10,0.00014) (11,0.0001475) (12,0.000156) (13,0.000186) (14,0.000197) (15,0.0002075) (16,0.000245) (17,0.000262) (18,0.000302) (19,0.0003145) (20,0.000351) (21,0.0004025) (22,0.000449) (23,0.0004645) (24,0.000511) (25,0.0006165) (26,0.0006195) (27,0.000724) (28,0.0007845) (29,0.000846) (30,0.0008975) (31,0.001045) (32,0.001108) (33,0.001187) (34,0.001327) (35,0.001423) (36,0.001538) (37,0.001639) (38,0.001819) (39,0.00192) (40,0.002047) (41,0.002168) (42,0.002354) (43,0.002541) (44,0.00268) (45,0.002759) (46,0.002982) (47,0.003087) (48,0.003289) (49,0.003482) (50,0.003544) (51,0.00372) (52,0.003942) (53,0.004093) (54,0.004165) (55,0.00439) (56,0.004457) (57,0.004696) (58,0.004801) (59,0.004929) (60,0.00517) (61,0.005298) (62,0.00536)};
\addplot+[color=rot, solid, line width=1.4pt, mark=none] coordinates {(0,8e-07) (1,1.8e-06) (2,2.85e-06) (3,4.05e-06) (4,5.8e-06) (5,7.35e-06) (6,1.01e-05) (7,1.36e-05) (8,1.875e-05) (9,2.13e-05) (10,2.815e-05) (11,3.13e-05) (12,4.135e-05) (13,4.71e-05) (14,5.47e-05) (15,6.36e-05) (16,8.105e-05) (17,9.31e-05) (18,0.0001063) (19,0.0001215) (20,0.0001422) (21,0.0001649) (22,0.0001912) (23,0.0002202) (24,0.0002532) (25,0.0002876) (26,0.0003213) (27,0.000368) (28,0.0004238) (29,0.0004761) (30,0.000549) (31,0.0006168) (32,0.0007022) (33,0.000802) (34,0.0009014) (35,0.001039) (36,0.001172) (37,0.00132) (38,0.001497) (39,0.001694) (40,0.001915) (41,0.00218) (42,0.002481) (43,0.002801) (44,0.003178) (45,0.003597) (46,0.004082) (47,0.004636) (48,0.005279) (49,0.006011) (50,0.006828) (51,0.007811) (52,0.008911) (53,0.01023) (54,0.01165) (55,0.01335) (56,0.01531) (57,0.01772) (58,0.02058) (59,0.02423) (60,0.02898) (61,0.0361) (62,0.05005)};
\draw[anthrazit, densely dashed, line width=0.7pt] (axis cs:30,1e-06) -- (axis cs:30,1);
\node[anthrazit, font=\scriptsize, anchor=west] at (axis cs:30,5.0e-06){\(k{=}30\)};

\addlegendimage{color=dunkelgrau, solid, line width=0.9pt, mark=none}
  \addlegendentry{\(\bm{L}_\mathrm{ch}\)}

  \addlegendimage{color=mittelblau, solid, line width=0.9pt, mark=none}
  \addlegendentry{BP-\(2(\bm{L}_\mathrm{ch})\)}

  \addlegendimage{color=apfelgruen, solid, line width=0.95pt, mark=none}
  \addlegendentry{BP-\(32(\bm{L}_\mathrm{ch})\)}
  
  \addlegendimage{color=rot, solid, line width=1.4pt, mark=none}
  \addlegendentry{\(\bm{L}_\mathrm{comb}\)}

\end{groupplot}
\node[
      anchor=north
  ] at (current bounding box.south) {\ref{ranklegendx}};

\end{tikzpicture}
\caption{Ordered-reliability \ac{BER} on \ac{BCH}(63,30). Sorted by \ac{LLR} magnitude, the soft-combined \ac{LLR} holds the error
rate low across the whole {MRB}. High iteration \ac{BP} distributes errors equally and increases the error rate in the high-reliability tail. For the calculation of $\Lcomb$: $M{=}16$ members, $I_\mathrm{max}{=}2$.}
\label{fig:rankber}
\end{figure}
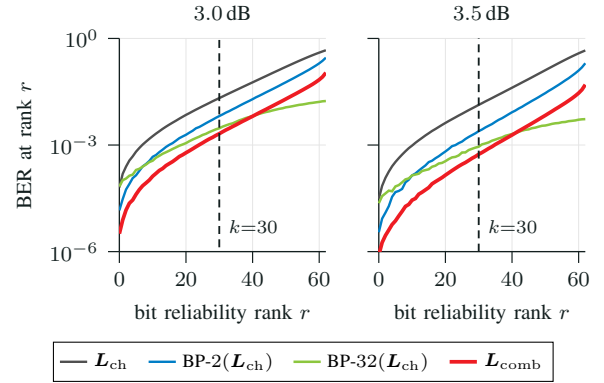

\Ac{URLLC} targets error probabilities as low as \(10^{-5}\) within a
\(1\,\mathrm{ms}\) user-plane latency budget~\cite{3gpp_tr38913}.
Such stringent latency and reliability requirements necessitate short
codes that can be decoded close to \ac{ML} performance.

For short, dense codes, iterative \ac{BP} decoding on a single graph
often saturates well before reaching \ac{ML} performance.
Ensemble decoding addresses this limitation by operating \(M\) decoders
in parallel on diverse representations of the same code and returning
the most likely valid candidate through an \ac{ML}-selection rule \cite{krieg2025comparative}.
Diversity stems, e.g., from multiple parity-check bases in
\ac{MBBP}~\cite{Hehn_MBBP_cyclic} or from automorphism-induced
permutations in \ac{AED}~\cite{geiselhart2022qc_aed}.

\begin{figure*}[!t]
\centering
\colorlet{hardgrey}{black!45}
\begin{tikzpicture}[
    font=\footnotesize,
    >={Triangle[angle'=60, length=1.9mm]},
    rbblock/.style={
        draw,
        line width=0.75pt,
        fill=#1!8,
        align=center,
        minimum height=8.5mm,
        minimum width=16mm,
        inner xsep=3pt,
        inner ysep=2pt
    },
    greyblock/.style={
        rbblock=white,
        draw=black!55,
        text=black!70
    },
    tallblock/.style={
        draw,
        fill=white,
        align=center,
        minimum height=42mm,
        minimum width=10mm,
        inner xsep=2pt,
        inner ysep=2pt
    },
    stagebox/.style={
        draw=#1,
        line width=1.5pt,
        fill=#1!8,
        rounded corners=2pt,
        inner sep=2.2mm
    },
    arr/.style={->, line width=0.55pt, draw=black},
    arrg/.style={->, line width=0.55pt, draw=black!55},
    line/.style={line width=0.55pt, draw=black},
    note/.style={font=\scriptsize, align=center}
]
\def\xIn{-0.4}       %
\def\xBusA{0.55}     %
\def\xColA{2.35}     %
\def\xColB{4.55}     %
\def\xComb{6.85}     %
\def\xColC{9.95}     %
\def\xDots{11.30}    %
\def\xColD{12.65}    %
\def\xML{14.70}      %
\def\xOut{16.40}     %
\def\xHoD{\xDots}    %
\def\yA{1.50}        %
\def\yB{0.15}        %
\def\yDots{-0.85}    %
\def\yC{-1.90}       %
\def\yMid{-0.20}     %
\def\yLow{-3.55}     %
\node (llr) at (\xIn,\yMid) {\(\bm{L}_{\mathrm{ch}}\)};
\foreach \m/\y in {1/\yA, 2/\yB, M/\yC}{
    \node[rbblock=white] (a\m) at (\xColA,\y) {\(\mathrm{Dec}_{\m}\)};
    \node[rbblock=white] (b\m) at (\xColB,\y) {\(\mathrm{Dec}_{\m}\)};
}
\node[note] at (\xColA,\yDots) {\(\vdots\)};
\node[note] at (\xColB,\yDots) {\(\vdots\)};
\foreach \m/\y in {1/\yA, 2/\yB, M/\yC}{
    \node[greyblock] (c\m) at (\xColC,\y) {\(\mathrm{Dec}_{\m}\)};
    \node[greyblock] (d\m) at (\xColD,\y) {\(\mathrm{Dec}_{\m}\)};
    \node[note, text=black!45] (dt\m) at (\xDots,\y) {\(\cdots\)};
}
\node[note, text=black!45] at (\xColC,\yDots) {\(\vdots\)};
\node[note, text=black!45] at (\xColD,\yDots) {\(\vdots\)};
\node[tallblock, draw=black!55, text=black!70] (ml) at (\xML,\yMid) {
    \rotatebox{90}{\begin{tabular}{c}ML in\\[-1pt]the list\end{tabular}}
};
\begin{scope}[on background layer]
    \node[stagebox=apfelgruen,
          fit={(a1)(aM)(b1)(bM)([xshift=-2mm]a1.west)}] (box1) {};
    \node[stagebox=hardgrey,
          fit={(c1)(cM)(d1)(dM)(ml)([xshift=-2mm]c1.west)}] (box2) {};
\end{scope}
\node[align=center, anchor=south] (lab1) at ($(box1.north)+(0,0.05)$)
    {\textcolor{apfelgruen}{Soft}-out\\\textcolor{apfelgruen}{ensemble} stage};
\node[black!45, align=center, anchor=south] at ($(box2.north)+(0,0.05)$)
    {Hard-out\\ ensemble stage};
\coordinate (busL) at (\xBusA,\yMid);
\draw[line] (llr.east) -- (busL);
\draw[line] (busL |- a1.west) -- (busL |- aM.west);
\foreach \m in {1, 2, M}{
    \draw[arr] (busL |- a\m.west) -- (a\m.west);
}
\foreach \m in {1, 2, M}{
    \draw[arr] (a\m) -- (b\m);
}
\foreach \m in {1, 2, M}{
    \draw[arr] (b\m.east) -- node[above, font=\scriptsize, pos=0.16]
        {\(\bm{e}_{\m}\)} (c\m.west);
}
\foreach \m in {1, 2, M}{
    \draw[arrg] (c\m) -- (dt\m);
    \draw[arrg] (dt\m) -- (d\m);
    \draw[arrg] (d\m.east) -- node[above, font=\scriptsize,
        text=black!45] {\(\hat{\bm{x}}_{\m}\)} (ml.west |- d\m.east);
}
\node (out) at (\xOut,\yMid) {\(\hat{\bm{c}}\in\mathcal{C}\)};
\draw[arr] (ml.east) -- (out.west);
\node[tallblock, draw=none, fill=mittelblau!20]
    (cand) at (\xComb,\yMid) {
    \rotatebox{90}{%
        \begin{tabular}{c}
            \textcolor{mittelblau}{combining stage}
        \end{tabular}%
    }
};
\node[tallblock, draw=none, minimum width=6mm, inner xsep=1pt,
      fill=orange!30, anchor=west]
    (post) at (cand.east) {\rotatebox{90}{post-processing}};
\draw[line width=1pt, densely dashed]
    (cand.north west) rectangle (post.south east);
\node[rbblock=mittelblau, draw=mittelblau, fill=mittelblau!20, line width=1pt]
    (hod) at (\xHoD,\yLow) {hard-out decoder};
\draw[arr, thick, mittelblau] (cand.south) |- node[pos=0.7, above, mittelblau] {$\Lm_\mathrm{comb}$} (hod.west);

\node[anchor=south, mittelblau] at (hod.north) {post-processing};

\node[inner sep=0,
      fit={([shift={(-1.4mm,1.2mm)}]lab1.north west)
           ([shift={(-1.4mm,-1.4mm)}]box1.south west)
           (cand.north east) (cand.south east)}] (secs) {};
\draw[mittelblau, line width=1.2pt, rounded corners=3pt]
    (cand.north east) -- (secs.north east) -- (secs.north west) --  (secs.south west) -- (secs.south east) -- (cand.south east);
\draw[black, line width=1pt, densely dashed]
    (cand.north east) -- (cand.south east);
\node[anchor=north, align=center] at ($(secs.south)+(0,-0.06)$)
    {\textcolor{apfelgruen}{SE}\textcolor{mittelblau}{CS}:
     \textcolor{apfelgruen}{soft ensemble}-\textcolor{mittelblau}{combining
     stage}};

\draw[arrg] (busL) -- ++(0,3.5) -| (ml);
\draw[arr] (busL) -- ++(0,3.5) -| (cand);
\draw[arr, mittelblau] (hod.east) -- node[pos=1.0, anchor=west, mittelblau]{$\hat{\bm{c}}\in\mathcal{C}$} ++(0.5,0);
\end{tikzpicture}
\caption{SECS-based decoding architecture.  \(M\) diverse \ac{BP}
members run for a few iterations only; the SECS
({\textcolor{mittelblau}{blue}} border) fuses their clipped extrinsics
into \(\Lcomb\).  The dashed unit is inserted into a conventional
hard-out ensemble decoder ({\textcolor{apfelgruen}{green}} and {\textcolor{black!55}{gray}}): \(\Lcomb\)
either feeds a re-encoding decoder
directly or re-enters the ensemble for
further ensemble-decoding after some post-processing ({\textcolor{orange}{orange}}).}
\label{fig:arch}
\vspace{-1em}
\end{figure*}
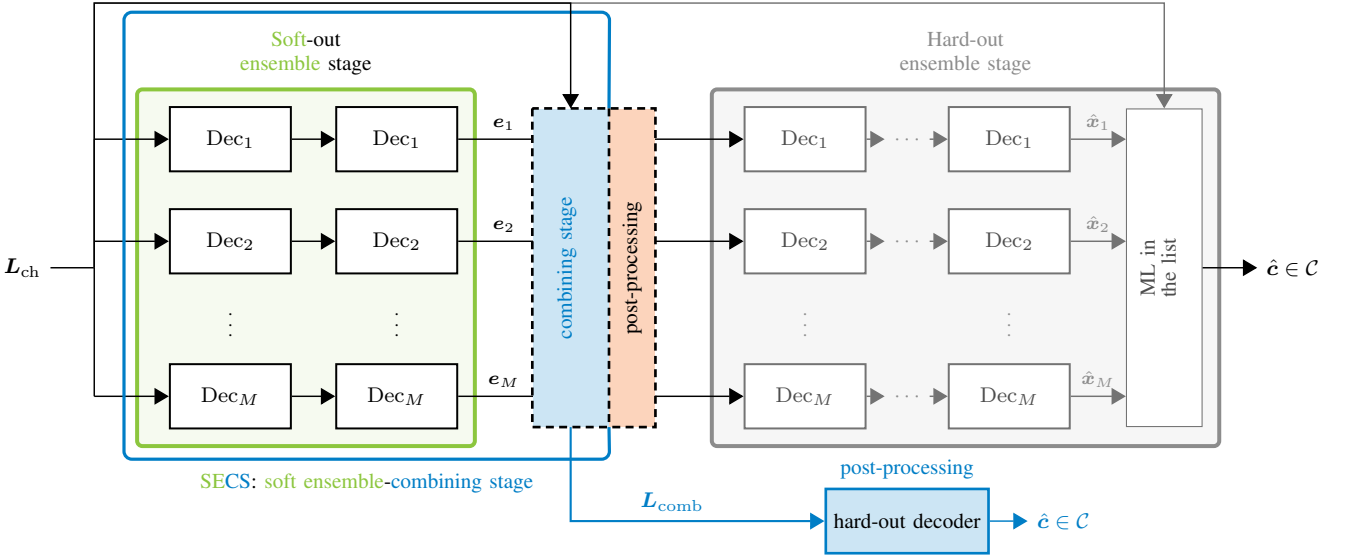

A conventional selection-only ensemble is bounded by its \ac{LER}, defined
as the probability that none of its constituent decoders produces the
transmitted codeword.
No selection rule can recover such a frame from the member candidate list.
Surpassing this bound therefore requires synthesizing an additional candidate
from information that is distributed across the ensemble members.

This paper starts from a simple observation:
shallow, diverse \ac{BP} members correct little, but they \emph{sort}
exceptionally well, as shown in Fig.~\ref{fig:rankber}.
Based on this observation, we propose the \ac{SECS}.
After only a few iterations, \ac{SECS} aggregates the clipped extrinsic
messages of the ensemble members into a combined \ac{LLR} vector whose
reliability ordering concentrates the residual errors in the least
reliable positions.
A re-encoding post-processing -- low-order \ac{OSD}~\cite{fossorier1995osd} or a
variant without \ac{GE} -- then recovers the
codeword from the nearly error-free most reliable positions.
In the studied configurations, deeper \ac{BP} decoding generally reduces
the average \ac{BER}, but also flattens the ordered-reliability error
profile. Consequently, its remaining errors can occur inside the
\ac{MRB}, making the posterior less suitable for re-encoding even when
its average \ac{BER} is lower.

Feeding \ac{BP} soft output to \ac{OSD} is a classical
idea~\cite{fossorier2001BPOSD}.
Prior work improves the \ac{OSD} input through temporal accumulation
and decoder diversity.
The authors of \cite{jiang2007likelihood_osd} accumulate \ac{LLR}
transitions across \ac{BP} iterations.
The approach of \cite{rosseel2023complementary_osd} learns such
temporal combinations and processes complementary reliability vectors
by separate \ac{OSD} instances.
An earlier work~\cite{rosseel2022diversity_osd} applies \ac{OSD}
separately to complementary, absorbing-set-specialized
\mbox{\ac{BP}-RNN} decoders.
\Ac{MBBP}-X~\cite{Hehn_MBBP_cyclic} combines soft
output across parity-check bases, but feeds it back into the
constituent decoders. 

In contrast, we fuse clipped extrinsic information \emph{across ensemble
members} at a common shallow iteration and apply a re-encoding
to the resulting reliability vector.
The proposed method therefore synthesizes a candidate from information
distributed across several decoding trajectories, rather than applying
post-processing separately to member-wise or iteration-wise soft outputs.
This distinction is particularly relevant for short, dense graphs, where
individual \ac{BP} posteriors frequently contain erroneous bits with
saturated reliability.
Shallow cross-member fusion suppresses such member-specific errors while
preserving the steep reliability ordering required by low-order
\ac{OSD}.

The main contributions of this paper are as follows:
\begin{itemize}
    \item We propose the \ac{SECS}, which combines the member
    extrinsics of a \ac{BP} ensemble after only a few iterations and
    passes the result to a re-encoding stage.
    The combining weight is optimized for each studied ensemble.
    \item We analyze the soft-combining gain via ordered-reliability
    error profiles and derive a closed-form \ac{FER} estimate from them.
    \item We compare the proposed \ac{SECS} to selection-based ensemble
    decoding on three codes and identify the regimes in which combining
    is beneficial.
    \item We describe a search procedure for cyclic codes with many
    low-weight dual codewords.
    Plain \ac{BP} fails to decode the resulting \((105,53)\) code,
    while \ac{OSD} with \ac{SECS} input operates close to \ac{ML}.
    \item We present an alternative post-processing decoder.
    It can be interpreted as an \ac{OSD} whose re-encoding stage is
    computed by \ac{BP} hardware, enabling \ac{OSD}-like performance
    where a full \ac{OSD} is not affordable.
\end{itemize}
Fig.~\ref{fig:arch} shows the resulting \ac{SECS}-based architecture.
\(M\) parallel \ac{BP} decoders run for a few iterations only.
Instead of continuing as separate decoders, the \ac{SECS} ({\textcolor{mittelblau}{blue}} border) aggregates their extrinsics.
The combined \acp{LLR} are then passed to a re-encoding stage (in {\textcolor{mittelblau}{blue}}).
Optionally, a post-processing step (in {\textcolor{orange}{orange}}) prepares the combined \acp{LLR} as
input for another round of classical ensemble decoding (in {\textcolor{black!55}{gray}}).
Alternatively, the \ac{BP} ensemble itself can be used to mimic the re-encoding step of an \ac{OSD}.

\section{Preliminaries and System Model}
\label{sec:prelims}

We consider a binary linear code \(\mathcal{C}\) of length \(n\) and
dimension \(k\), represented by a \ac{PCM} $\mathbf{H}$ fulfilling $\bm{c}\mathbf{H}^\top=\bm{0}$ with codeword \(\bm{c}\in\{0,1\}^n\).
Codewords are modulated using \ac{BPSK} as
\(\bm{x} = 1-2\bm{c}\) and transmitted over the \ac{AWGN} channel, i.e.,
\(\bm{y} = \bm{x} + \bm{n}\) with
\(\bm{n}\sim\mathcal{N}(\bm{0},\sigma^2\bm{I})\).
The channel \ac{LLR} vector is
\begin{equation}
\Lch = \frac{2}{\sigma^2}\,\bm{y}.
\label{eq:llr}
\end{equation}

An ensemble decoder consists of \(M\) constituent \ac{BP} decoders for
\(\mathcal{C}\).
In this work, diversity among the members stems from two sources.
In \emph{multiple-bases} ensembles~\cite{Hehn_MBBP_cyclic}, the members
use structurally different sparse \acp{PCM} of the same code.
For the \ac{BCH} codes, we employ \ac{ssPCM}
constructions~\cite{shen2025universal,shen2025sspcm}, in which each dense dual row is
decomposed into several low-weight checks, coupled through punctured
\acp{AVN}.
For codes with redundant duals, we draw independent rank-preserving row
subsets of an overcomplete \ac{PCM}.
In \emph{automorphism ensembles}~\cite{geiselhart2022qc_aed},
a single matrix is used under coordinate permutations from the
automorphism group of the code.
The finite-geometry codes considered below combine both views: their
duals contain the incidence vectors of all lines of the underlying
geometry~\cite{kou2001finite}.
These form an overcomplete \ac{PCM} that can be sparsified and
permuted~\cite{shen2026lifted}.

Each ensemble member \(m\) runs flooding \ac{BP} and stops after
\(I_\mathrm{max}\) iterations.
Its extrinsic message vector on the \(n\) code bits is \(\bm{e}_m\).
For \ac{ssPCM} members, the \ac{SECS} combines the posteriors of the
\(n\) code-bit variable nodes only; the punctured \acp{AVN}
(initialized to zero, without channel \acp{LLR}) are never combined, and
\(\bm{e}_m\) is the code-bit posterior minus \(\Lch\).
Single-iteration members employ one round of the \ac{NMSA}, whereas members with multiple iterations use \ac{SPA} updates~\cite{gallager1963ldpc,chen2002minsum}.
Both are attenuated by a factor \(\alpha\). \Ac{NMSA} uses \(\alpha=0.75\), \ac{SPA} uses \(\alpha=0.5\).
Single-iteration members use \ac{NMSA}.
Its suboptimal approximation cannot propagate within one iteration.
Multi-iteration members, in contrast, use the exact \ac{SPA} update,
damped for stability on the loopy, dense graphs.

As a reference post-processing, we use
\mbox{\ac{OSD}-1}~\cite{fossorier1995osd}: the input positions are
sorted by reliability, the \ac{MRB} is obtained by \ac{GE}, and the
decision is the highest-correlation candidate among the re-encoded
words of order at most one, i.e., all single-bit flips on the \ac{MRB}.
The distinction between the \(k\) most reliable positions and
the true \ac{MRB} matters: \ac{GE} needs to replace linearly dependent positions by less reliable ones.

\section{The Soft Ensemble-Combining Stage}
\label{sec:softcomb}

\begin{figure}[!t]
\centering
\begin{tikzpicture}
\begin{axis}[
    width=0.6\linewidth,
    height=0.6\linewidth,
    xmode=log, ymode=log, grid=both,
    log basis y={10},
    log basis x={2},
    log origin=infty,
    major grid style={draw=hellgrau!85, line width=0.36pt},
    minor grid style={draw=hellgrau!45, line width=0.22pt},
    axis x line*=bottom, axis y line*=left,
    axis line style={draw=anthrazit, line width=0.50pt},
    tick align=outside, tick style={draw=anthrazit, line width=0.45pt},
    ticklabel style={font=\footnotesize, text=anthrazit},
    label style={font=\footnotesize, text=anthrazit},
    xtick={0.5,1,2,4,8,16},
    xticklabels={$0.5$,$1$,$2$,$4$,$8$,$16$},
    xmin=0.5, xmax=16,
    ymin=5e-05, ymax=1,
    xlabel={soft-combining weight \(wM\)},
    ylabel={FER}
]

  \addplot+[color=anthrazit, mark=*, mark size=1.25pt, solid, line width=0.9pt, mark options={solid}] coordinates {(0.5,0.7393) (1,0.5126) (1.5,0.3807) (2,0.2987) (3,0.2092) (4,0.1685) (6,0.1399) (8,0.1366) (12,0.1438) (16,0.1532)};
  \label{plot:hard3}
  \addplot+[color=mittelblau, mark=*, mark size=1.25pt, solid, line width=0.9pt, mark options={solid}] coordinates {(0.5,0.5577) (1,0.3089) (1.5,0.1998) (2,0.1434) (3,0.09311) (4,0.07414) (6,0.06433) (8,0.06572) (12,0.07224) (16,0.07875)};
  \label{plot:hard35}
  \addplot+[color=apfelgruen, mark=*, mark size=1.25pt, solid, line width=0.9pt, mark options={solid}] coordinates {(0.5,0.3779) (1,0.1516) (1.5,0.08499) (2,0.05553) (3,0.03347) (4,0.02648) (6,0.02435) (8,0.02605) (12,0.03) (16,0.03298)};
  \label{plot:hard4}
  \addplot+[color=anthrazit, mark=*, mark size=1.25pt, dotted, line width=0.9pt, mark options={solid}] coordinates {(0.5,0.00718) (1,0.00508) (1.5,0.00414) (2,0.00362) (3,0.00328) (4,0.00331) (6,0.00377) (8,0.00474) (12,0.00889) (16,0.01313)};
  \label{plot:osd3}
  \addplot+[color=mittelblau, mark=*, mark size=1.25pt, dotted, line width=0.9pt, mark options={solid}] coordinates {(0.5,0.00175) (1,0.00103) (1.5,0.00077) (2,0.00066) (3,0.00062) (4,0.00065) (6,0.00086) (8,0.0014) (12,0.00306) (16,0.00487)};
  \label{plot:osd35}
  \addplot+[color=apfelgruen, mark=*, mark size=1.25pt, dotted, line width=0.9pt, mark options={solid}] coordinates {(0.5,0.00036) (1,0.0002) (1.5,0.00012) (2,9e-05) (3,0.00012) (4,0.00014) (6,0.00026) (8,0.00042) (12,0.00082) (16,0.00136)};
  \label{plot:osd4}
\end{axis}
\node (tab) [
      right = 0.12cm of current bounding box.east,
      shape=rectangle,
      anchor=west,
      font=\scriptsize,
      fill=white,
      draw=black,
      inner xsep=3pt,
      inner ysep=4pt,
  ] {
      \begin{NiceTabular}{lc}
          hard dec.
          &
          \tikz[baseline=-0.6ex]{\draw[anthrazit, solid, line width=0.9pt] (0,0) -- (0.38,0);}
          \\
          OSD-1
          &
          \tikz[baseline=-0.6ex]{\draw[anthrazit, dotted, line width=0.9pt] (0,0) -- (0.38,0);}
          \\
          \(3\,\mathrm{dB}\)
          &
          \tikz[baseline=-0.6ex]{
              \draw[anthrazit, solid, line width=0.9pt] (0,0) -- (0.38,0);
              \fill[anthrazit] (0.19,0) circle[radius=2pt];
          }
          \\
          \(3.5\,\mathrm{dB}\)
          &
          \tikz[baseline=-0.6ex]{
              \draw[mittelblau, solid, line width=0.9pt] (0,0) -- (0.38,0);
              \fill[mittelblau] (0.19,0) circle[radius=2pt];
          }
          \\
          \(4\,\mathrm{dB}\)
          &
          \tikz[baseline=-0.6ex]{
              \draw[apfelgruen, solid, line width=0.9pt] (0,0) -- (0.38,0);
              \fill[apfelgruen] (0.19,0) circle[radius=2pt];
          }
          \\
      \end{NiceTabular}
  };
\end{tikzpicture}
\caption{Soft-combining weight sweep for the BCH\((63,30)\) ssPCM ensemble.
  After $I_\mathrm{max}=2$ iterations of an \(M=16\) ensemble, the extrinsics are combined according to Eq.~\eqref{eq:comb} and decoded by either hard decision or \mbox{\ac{OSD}-1}.}
\label{fig:msweep}
\end{figure}
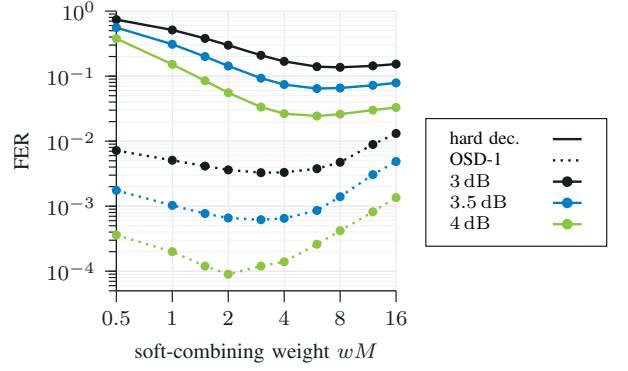

Instead of selecting one output after all members have converged, the
\ac{SECS} combines the ensemble into a single soft word.
The members stop after \(I_\mathrm{max}\in\{1,2\}\) iterations and their clipped extrinsics are accumulated onto the channel \ac{LLR} $\Lch$,
yielding the combined \ac{LLR}
\begin{equation}
\Lcomb = \Lch + w \sum_{m=1}^{M} \clip\!\big(\bm{e}_m,\pm\delta\big),
\label{eq:comb}
\end{equation}
where \(\clip(\cdot)\) limits its input to \(\pm\delta\) and we set
\(\delta=8\).
This hyperparameter appears very robust: varying \(\delta\) over
\(\{4,8,16,64\}\) changes the measured \mbox{\ac{OSD}-1} \ac{FER} at
the operating point of Fig.~\ref{fig:msweep} by less than
\(2\,\%\).

We state the combiner weight in the form \(w\!\cdot\!M\).
Then, \(wM{=}1\) corresponds to the mean of the extrinsics, and
\(wM{=}M\) to their sum, which is optimal for uncorrelated observations.
In all studied ensembles, the optimal \(wM\) also compensates the
attenuation \(\alpha\) of the tapped extrinsics.
Members with deep \ac{AVN} trees, read at an early tap, require
\(wM{\approx}M\).
Shallow structures require \(wM{\approx}2\)--\(4\), and full-strength
single-iteration members \(wM{=}1\).
If the amplification is too strong, the residual member errors are
pushed back into the reliable positions, and the ordering degrades.
Fig.~\ref{fig:msweep} shows the sweep for the \ac{BCH} \((63,30)\)
ensemble.
The hard-decision optimum lies at moderate weights, whereas the
\ac{OSD} input prefers smaller ones.
This reflects a tradeoff: the hard decision must rescue all bits, which
sometimes harms the \ac{MRB}, whereas the \ac{OSD} input only requires
a good \ac{MRB}.

\begin{figure}[!t]
\centering
\input{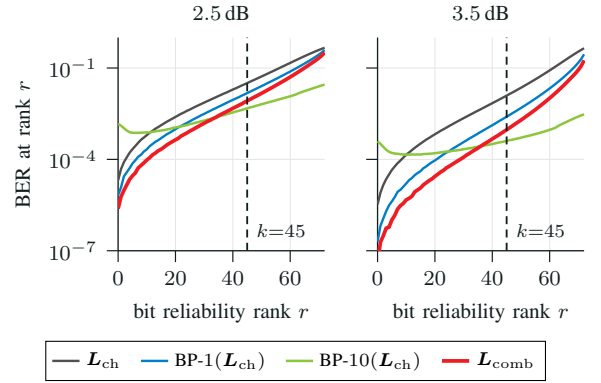}
\vspace{-1em}
\caption{Ordered-reliability \ac{BER} on PG(2,8) $(73,45)$: the combined one-iteration
ensemble produces the lowest \ac{BER} in the \ac{MRB}. The ten-iteration full-\ac{PCM} decoder distributes \ac{BER} evenly and shows an overconfidence plateau at its most reliable positions. For $\Lcomb$: $M{=}8$, $I_\mathrm{max}{=}1$.}
\label{fig:rankber_long}
\vspace{-1em}
\end{figure}

To quantify what the combining achieves, we sort the positions of a
soft word by their reliability \(|L_i|\) and measure the error rate at
each rank \(r\).
Averaged over many frames, this yields the ordered-reliability profile
\(\{b_r\}\) of the decoder input.
Fig.~\ref{fig:rankber} shows these profiles for a \ac{BCH} \((63,30)\)
\ac{ssPCM} ensemble.
The combined \ac{LLR} holds the error rate approx. below \(10^{-3}\) across the
entire \ac{MRB} region, one to two orders of magnitude below the raw
channel.
Moreover, it preserves a steep reliability ordering of the channel
observation and concentrates the residual errors in the least reliable
bits.
In contrast, \ac{BP} primarily minimizes the average
bit-error probability and thereby produces a flatter
ordered-reliability profile after some iterations.
For a re-encoding decoder however, errors do not need to be corrected directly as
long as they are reliably sorted out of the information set.
Fig.~\ref{fig:rankber_long} shows the corresponding profiles for
\(\mathrm{PG}(2,8)\).

In the \(45\) most reliable positions, the combination of eight
one-iteration members runs below the posterior of a ten-iteration
decoder on the full overcomplete matrix. Since this code uses no \ac{ssPCM} and, thus, no \acp{AVN}, a single iteration suffices to generate useful soft information.
The deep decoder is better only in the tail, which the \ac{OSD} post-processing
discards.
Moreover, the deep profile is non-monotonic at the top positions: its most
confident bits are wrong more often than its rank-\(20\) bits.
This is caused by occasional convergence to a wrong codeword at
saturated confidence.
The one-iteration combination is free of this failure mode, since no
member has converged.
Iterated \ac{BP} redistributes reliability from strong positions to
weak ones and thereby lowers the overall \ac{BER}.
For a stand-alone decoder, this is the right objective.
With a re-encoding stage attached, however, only the ordering matters,
and the redistribution is counterproductive.

The profiles also admit a simple \ac{FER} estimate.
For a set \(\mathcal{R}\) of positions,
\(\Sigma_{\mathcal{R}} = \sum_{r\in\mathcal{R}} b_r\) is the expected
number of errors among those positions.
Treating error events as independent across positions yields
\begin{equation}
\hat{P}_{\mathcal{R}} = 1 - e^{-\Sigma_{\mathcal{R}}}
\approx \Pr\{\text{at least one error in } \mathcal{R}\},
\label{eq:ferest}
\end{equation}
equivalent to modeling the error count in $\mathcal{R}$ as Poisson with
mean $\Sigma_{\mathcal{R}}$.
For \(\mathcal{R}=\{1,\dots,k\}\), i.e., the \(k\) most reliable positions
in Figs.~\ref{fig:rankber} and~\ref{fig:rankber_long},
\eqref{eq:ferest} estimates the input-failure rate of re-encoding
stages.
At the same time, it approximates the \mbox{\ac{OSD}-0} error rate,
since \mbox{\ac{OSD}-0} fails exactly when the information set contains
an error.

For \(\mathcal{R}=\{1,\dots,n\}\), \eqref{eq:ferest} estimates the
hard-decision \ac{FER}.
For the channel, the approximation is exact, as the noise -- and therefore the error
positions -- are independent. For correlated decoder outputs, the approximation is a less tight upper bound.

\begin{figure}[t]
      \centering
      \input{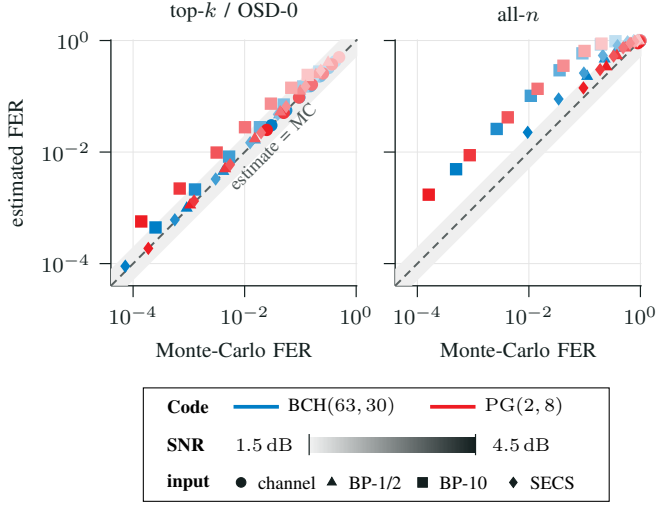}
      \vspace{-1em}
      \caption{Estimated versus simulated \ac{FER} obtained from sorted-\ac{BER}
      profiles. Colors distinguish the codes, color saturation represents the
      \ac{SNR}, and markers identify the decoder input. The dashed diagonal indicates
      exact agreement, while the gray region marks agreement within a factor of two.}
      \label{fig:FER_approx}
      \vspace{-1em}
  \end{figure}

Fig.~\ref{fig:FER_approx} compares the estimates against Monte-Carlo
simulation for profiles like the ones shown in Figs.~\ref{fig:rankber}
and~\ref{fig:rankber_long}.
Channel and \ac{SECS} inputs are approximated with small errors.
Deep-\ac{BP} posteriors, in contrast, violate the independence
assumption by up to an order of magnitude -- their failures arrive as
clusters of confident errors, the same mechanism as the overconfidence
plateau in Fig.~\ref{fig:rankber_long} -- which lead to worse approximation.

\section{Numerical Results}
\label{sec:results}
\begin{figure*}[!t]
\centering
\subfloat[\ac{BCH}(63,30) \ac{ssPCM} ensemble, $M{=}16$,
$I_\mathrm{max}{=}2$, $wM{=}1.5$.\label{fig:fer_bch63}]{%
\begin{tikzpicture}
\begin{axis}[
    scale only axis,
    width=50mm,
    height=40mm,
    ymode=log,
    log origin=infty,
    grid=both,
    major grid style={draw=hellgrau!85, line width=0.36pt},
    minor grid style={draw=hellgrau!45, line width=0.22pt},
    axis x line*=bottom,
    axis y line*=left,
    axis line style={draw=anthrazit, line width=0.50pt},
    tick align=outside,
    tick style={draw=anthrazit, line width=0.45pt},
    minor tick style={draw=anthrazit!55, line width=0.30pt},
    ticklabel style={font=\footnotesize, text=anthrazit},
    label style={font=\footnotesize, text=anthrazit},
    xmin=1, xmax=7,
    ymin=1e-05, ymax=1,
    xlabel={\(E_\mathrm{b}/N_0\) in dB},
    ylabel={FER},
    yticklabels={,$10^{-5}$,$10^{-4}$,$10^{-3}$,$10^{-2}$,$10^{-1}$,},
]

  \addplot+[color=anthrazit, mark=*, mark size=1pt, dashed, line width=0.9pt, mark options={solid}] coordinates {(0,0.91118421) (0.5,0.79605263) (1,0.61931818) (1.5,0.36993243) (2,0.23648649) (2.5,0.12607759) (3,0.052153716) (3.5,0.015438988) (4,0.003813244) (4.5,0.00093005952) (5,0.00015725644) (5.5,2.1855428e-05) (6.0, 3.520333e-06)};
  \addlegendentry{BP-10}

  \addplot+[color=anthrazit, mark=*, mark size=1pt, solid, line width=0.9pt, mark options={solid,fill=white}] coordinates {(0,0.57552083) (0.5,0.42773438) (1,0.3375) (1.5,0.20214844) (2,0.11997768) (2.5,0.069972826) (3,0.029775943) (3.5,0.010196834) (4,0.0024880573) (4.5, 0.0005946745562130177) (5,9.231479e-05) (5.5,1.0876067e-05)};
  \addlegendentry{BP-10 + OSD-1}
  
  \addplot+[color=R12, mark=square*, mark size=1pt, solid, line width=0.9pt, mark options={solid}] coordinates {(0,0.87369792) (0.5,0.71028646) (1,0.49609375) (1.5,0.28548177) (2,0.13802083) (2.5,0.04695638) (3,0.015104167) (3.5,0.0030790441) (4,0.0004733333) (4.5,5.6111e-5) (5.0, 4.3043e-6)};
  \addlegendentry{MBBP-16 BP-10}

  \addplot+[color=darkgreen, mark=square*, mark size=1pt, solid, line width=0.9pt, mark options={solid}] coordinates {(0,0.953125) (0.5,0.91015625) (1,0.85546875) (1.5,0.73697917) (2,0.671875) (2.5,0.45703125) (3,0.3390625) (3.5,0.16796875) (4,0.082236842) (4.5,0.031700721) (5,0.010037579) (5.5,0.0025327621)
  (6,0.00037924757) (6.5, 4.514232e-05) (6.8, 7.604852e-06)};
  \addlegendentry{MBBP-16 BP-2}

  \addplot+[color=mittelblau, mark=diamond*, mark size=2pt, dashed, line width=0.9pt, mark options={solid}] coordinates {(0,0.921875) (0.5,0.83984375) (1,0.76953125) (1.5,0.65820312) (2,0.53515625) (2.5,0.36197917) (3,0.24414062) (3.5,0.12388393) (4,0.049804688) (4.5,0.01732337) (5,0.0045955882) (5.5,0.00081550104)
  (6,9.1044167e-05) (6.5, 8.127913e-06) };
  \addlegendentry{\shortstack[l]{MBBP-16 BP-2\\soft comb.}}

  \addplot+[color=mittelblau, mark=diamond*, mark size=2pt, solid, line width=1.2pt, mark options={solid, fill=white}] coordinates {(0,0.38541667) (0.5,0.25683594) (1,0.1765625) (1.5,0.0859375) (2,0.037574405) (2.5,0.012807377) (3,0.0038675743) (3.5,0.0009539072) (4,0.00012765523) (4.5,1.8036e-05 )};
  \addlegendentry{\shortstack[l]{MBBP-16 BP-2 soft comb.\\+ OSD-1 [proposed]}}

  \addplot+[color=anthrazit, mark=none, dotted, line width=1.0pt]
  coordinates {
      (0,0.3215)
      (0.5,0.1969)
      (1,0.1161)
      (1.5,0.05631)
      (2,0.02315)
      (2.5,0.007628)
      (3,0.0022)
      (3.5,0.0004896)
      (4,7.973e-05)
      (4.5,8.396e-06)
  };
  \addlegendentry{ML}

\legend{}

\end{axis}
\end{tikzpicture}}
\hfill
\subfloat[PG(2,8) $(73,45)$, \ac{AED} ensemble, $M{=}8$, $I_\mathrm{max}{=}1$,
$wM{=}1$.\label{fig:fer_pg28}]{%
\begin{tikzpicture}[
spy using outlines={%
rectangle,%
magnification=2.5,%
width=20mm,%
height=18mm,%
every spy on node/.style={%
draw=gray!50,
line width=0.75pt,
fill=none
},%
every spy in node/.style={%
draw=black,
line width=0.75pt,
fill=white
},%
spy connection path={%
  \path
    (tikzspyonnode.north east) coordinate (onNE)
    (tikzspyonnode.north west) coordinate (onNW)
    (tikzspyonnode.south east) coordinate (onSE)
    (tikzspyonnode.south west) coordinate (onSW)
    (tikzspyinnode.north west) coordinate (inNW)
    (tikzspyinnode.south west) coordinate (inSW)
    (tikzspyinnode.north east) coordinate (inNE)
    (tikzspyinnode.south east) coordinate (inSE);%
  \draw[gray!50, solid,line width=0.55pt]
    (onNE) --
    (intersection of inNW--inSW and inNE--onNE);%
  \draw[gray!50, solid,line width=0.55pt]
    (inSW) --
    (onSW);%
  \draw[gray!50, solid,line width=0.55pt]
  (inSE) --(onSE);%
  \draw[gray!50, solid,line width=0.55pt]
  (inNW) --(onNW);%
}%
}%
]

\begin{axis}[
    scale only axis,
    width=50mm,
    height=40mm,
    ymode=log,
    log origin=infty,
    grid=both,
    major grid style={draw=hellgrau!85, line width=0.36pt},
    minor grid style={draw=hellgrau!45, line width=0.22pt},
    axis x line*=bottom,
    axis y line*=left,
    axis line style={draw=anthrazit, line width=0.50pt},
    tick align=outside,
    tick style={draw=anthrazit, line width=0.45pt},
    minor tick style={draw=anthrazit!55, line width=0.30pt},
    ticklabel style={font=\footnotesize, text=anthrazit},
    yticklabel=\empty,
    label style={font=\footnotesize, text=anthrazit},
    legend cell align={left},
    legend image code/.code={
        \draw[mark repeat=2, mark phase=2, #1]
            plot coordinates {(0cm,0cm) (0.16cm,0cm) (0.32cm,0cm)};
    },
    legend style={
        draw=black,
        fill=white,
        at={(0.46,-0.275)},
        anchor=north,
        legend columns=3,
        font=\scriptsize,
        inner xsep=1.5pt,
        /tikz/every even column/.append style={column sep=0.08cm}
    },
    xmin=1, xmax=7,
    ymin=1e-5, ymax=1,
    xlabel={\(E_\mathrm{b}/N_0\) in dB},
  name=mainplot,
]

  \addplot+[color=anthrazit, mark=*, mark size=1pt, dashed, line width=0.9pt, mark options={solid}] coordinates {(1,0.41503906) (1.5,0.24674479) (2,0.13736979) (2.5,0.059709821) (3,0.023129112) (3.5,0.0060579979) (4,0.0014141832) (4.5,0.00020125105574324325) (5.0, 2.756e-05)};
  \addlegendentry{BP-10 full OC}

  \addplot+[color=anthrazit, mark=*, mark size=1pt, solid, line width=0.9pt, mark options={solid, fill=white}] coordinates {(1,0.3671875) (1.5,0.22070312) (2,0.12044271) (2.5,0.052176339) (3,0.01953125) (3.5,0.0050428143) (4,0.0011727373) (4.5,0.00016495988175675675) (5.0, 2.455e-05)};
  \addlegendentry{BP-10 full OC + OSD-1}

  \addplot+[color=R12, mark=square*, mark size=1pt, solid, line width=0.9pt, mark options={solid}] coordinates {(1,0.41796875) (1.5,0.24023438) (2,0.12890625) (2.5,0.053955078) (3,0.01625) (3.5,0.0042459239) (4,0.00072742086) (4.5, 1.2087e-4)};
  \addlegendentry{AED-8 BP-10}
  
  \addplot+[color=darkgreen, mark=square*, mark size=1pt, solid, line width=0.9pt, mark options={solid}] coordinates {(1,0.9453125) (1.5,0.86425781) (2,0.734375) (2.5,0.59375) (3,0.4375) (3.5,0.27570907) (4,0.14516411) (4.5,0.061889648) (5,0.022094727) (5.5,0.0062866211) (6,0.0012032645) (6.5,0.00021151839) (7,2.4414063e-05)};
  \addlegendentry{AED-8 BP-1}

  \addplot+[color=mittelblau, mark=diamond*, mark size=2pt, dashed, line width=0.85pt, mark options={solid}] coordinates {(1,0.95898438) (1.5,0.88476562) (2,0.77001953) (2.5,0.63989258) (3,0.49304687) (3.5,0.33319888) (4,0.19335938) (4.5,0.090087891) (5,0.04309082) (5.5,0.018798828) (6,0.0067836217) (6.5,0.0034821609) (7,0.0022052376) (7.5, 0.0013941714638157894) (8.0,0.00114857)};
  \addlegendentry{\shortstack[l]{AED-8 BP-1\\soft comb.}}
  
  \addplot+[color=mittelblau, mark=diamond*, mark size=2pt, solid, line width=0.95pt, mark options={solid,fill=white}] coordinates {(1,0.31640625) (1.5,0.15429688) (2,0.077799479) (2.5,0.032226562) (3,0.011050576) (3.5,0.0024386476) (4,0.00045917839) (4.5,6.4079e-5)};
  \addlegendentry{\shortstack[l]{AED-8 BP-1 soft comb.\\+ OSD-1 [proposed]}}

  \addplot+[color=anthrazit, mark=none, dotted, line width=1.0pt] coordinates {(1,0.30371094) (1.5,0.13932292) (2,0.074869792) (2.5,0.028738839) (3,0.010279605) (3.5,0.0022289901) (4,0.00043115342) (4.5, 5.889e-5)};
  \addlegendentry{OSD-3}

\coordinate (zoomSW)     at (axis cs:3,1e-3);
\coordinate (zoomNE)     at (axis cs:5,1e-3);
\coordinate (zoomCenter) at (axis cs:3.9,1e-3);
\legend{}
\end{axis}

\spy
  on (zoomCenter)
  in node[anchor=north east]
  at ([xshift=-0.12cm,yshift=-0.12cm]mainplot.north east);

\end{tikzpicture}}
\hfill
\subfloat[Designed $(105,53)$ cyclic code,
\ac{MBBP} ensemble, $M{=}8$, $I_\mathrm{max}{=}2$,
$wM{=}1$.\label{fig:fer_c93}]{%
\begin{tikzpicture}
\begin{axis}[
    scale only axis,
    width=50mm,
    height=40mm,
    ymode=log,
    log origin=infty,
    grid=both,
    major grid style={draw=hellgrau!85, line width=0.36pt},
    minor grid style={draw=hellgrau!45, line width=0.22pt},
    axis x line*=bottom,
    axis y line*=left,
    axis line style={draw=anthrazit, line width=0.50pt},
    tick align=outside,
    tick style={draw=anthrazit, line width=0.45pt},
    yticklabel=\empty,
    minor tick style={draw=anthrazit!55, line width=0.30pt},
    ticklabel style={font=\footnotesize, text=anthrazit},
    label style={font=\footnotesize, text=anthrazit},
    legend cell align={left},
    legend image code/.code={
        \draw[mark repeat=2, mark phase=2, #1]
            plot coordinates {(0cm,0cm) (0.16cm,0cm) (0.32cm,0cm)};
    },
    legend style={
        draw=black,
        fill=white,
        at={(0.46,-0.275)},
        anchor=north,
        legend columns=3,
        font=\scriptsize,
        inner xsep=1.5pt,
        /tikz/every even column/.append style={column sep=0.08cm}
    },
    xmin=1, xmax=7,
    ymin=1e-05, ymax=1,
    xlabel={\(E_\mathrm{b}/N_0\) in dB},
]

  \addplot+[color=mittelblau, mark=diamond*, mark size=2pt, dashed, line width=0.9pt, mark options={solid}] coordinates {(0,1.0) (0.5,0.9999) (1,0.9999) (1.5,0.9979) (2,0.993) (2.5,0.9734) (3,0.9266) (3.5,0.8324) (4,0.6832) (4.5,0.5009) (5,0.3222) (5.5,0.1817) (6,0.09485) (6.5,0.04415) (7,0.0168)};
  \addlegendentry{\shortstack[l]{MBBP-8 BP-1\\soft comb.}}

  \addplot+[color=mittelblau, mark=diamond*, mark size=2pt, solid, line width=1.2pt, mark options={solid, fill=white}] coordinates {(0,0.665) (0.5,0.5041) (1,0.3518) (1.5,0.1998) (2,0.095) (2.5,0.04055) (3,0.01355) (3.5,0.003133) (4,0.00062) (4.5,0.0001233) (5,1.6e-05)};
  \addlegendentry{\shortstack[l]{MBBP-8 BP-1 soft comb.\\+ OSD-1 [proposed]}}
  \label{plot:SECS}

  \addplot+[color=dunkelgrau, mark=x, mark size=2pt, dash dot, line width=0.9pt, mark options={solid}] coordinates {(0,0.7148) (0.5,0.5626) (1,0.4156) (1.5,0.2723) (2,0.16) (2.5,0.0809) (3,0.03487) (3.5,0.01418) (4,0.00482) (4.5,0.001284) (5,0.0003225) (5.5,6.85e-05) (6,1.65e-05)};
  \addlegendentry{channel OSD-1}

  \addplot+[color=dunkelgrau, mark=none, dotted, line width=1.0pt] coordinates {(0,0.5673) (0.5,0.3912) (1,0.2431) (1.5,0.1253) (2,0.0526) (2.5,0.01873) (3,0.005133) (3.5,0.001183) (4,0.00026) (4.5,3.4e-05) (5,3e-06)};
  \addlegendentry{OSD-3}

  \addplot+[color=orange, mark=square*, mark size=1.0pt, solid, line width=0.9pt, mark options={solid}] coordinates {(0,1.0) (0.5,1.0) (1,1.0) (1.5,0.9991) (2,0.9909) (2.5,0.9494) (3,0.8168) (3.5,0.5538) (4,0.2723) (4.5,0.09472) (5,0.02074) (5.5,0.003057) (6,0.00028)};
  \addlegendentry{\shortstack[l]{MBBP-8 BP-10 \\ (BKLC~\cite{GrasslCodeTables} ssPCM)}}

  \addplot+[color=orange, mark=none, mark size=1.6pt, dotted, line width=0.9pt, mark options={solid}] coordinates {(0,0.4245) (0.5,0.2622) (1,0.1234) (1.5,0.0459) (2,0.0156) (2.5,0.00275) (3,0.000525) (3.5,4.783e-05) (4,2.7e-06)};
  \addlegendentry{OSD-3 (BKLC~\cite{GrasslCodeTables})}

  \addplot+[color=orange, mark=x, mark size=2pt, dash dot, line width=0.9pt, mark options={solid}] coordinates {(0,0.6564) (0.5,0.5187) (1,0.3642) (1.5,0.236) (2,0.1407) (2.5,0.06735) (3,0.03075) (3.5,0.01137) (4,0.00376) (4.5,0.00107) (5,0.000255) (5.5,7e-05) (6,9e-06)};
  \addlegendentry{channel OSD-1 (BKLC~\cite{GrasslCodeTables})}

  \addplot+[color=orange, mark=diamond*, mark size=2pt, solid, line width=1.2pt, mark options={solid, fill=white}] coordinates {(0,0.6573) (0.5,0.507) (1,0.3612) (1.5,0.2321) (2,0.1239) (2.5,0.0582) (3,0.023) (3.5,0.00702) (4,0.0015) (4.5,0.00026) (5,2.8e-05)};
  \addlegendentry{\shortstack[l]{SECS + OSD-1\\(BKLC~\cite{GrasslCodeTables} ssPCM)}}

\legend{}
\end{axis}
\end{tikzpicture}}
\\[0.5ex]
\begin{tikzpicture}
\begin{axis}[
    hide axis,
    scale only axis,
    width=1mm, height=1mm,
    xmin=0, xmax=1, ymin=0, ymax=1,
    legend cell align={left},
    legend image code/.code={
        \draw[mark repeat=2, mark phase=2, #1]
            plot coordinates {(0cm,0cm) (0.16cm,0cm) (0.32cm,0cm)};
    },
    legend style={
        draw=black,
        fill=white,
        at={(0.5,0.5)},
        anchor=center,
        legend columns=5,
        font=\scriptsize,
        inner xsep=1.5pt,
        /tikz/every even column/.append style={column sep=0.22cm}
    },
]
\addlegendimage{empty legend}
\addlegendentry{}
\addlegendimage{
    color=anthrazit,
    dashed,
    dash pattern=on 2pt off 1pt,
    dash phase=1.75pt,
    line width=0.9pt,
    mark=*,
    mark size=1pt,
    mark options={solid}
}
\addlegendentry{BP-10}
\addlegendimage{color=R12, mark=square*, mark size=1pt, solid, line width=0.9pt, mark options={solid}}
\addlegendentry{ensemble-$M$ BP-10}
\addlegendimage{color=mittelblau, mark=diamond*, mark size=2pt, dashed,
    dash pattern=on 2pt off 1pt,
    dash phase=1.75pt, line width=0.9pt, mark options={solid}}
\addlegendentry{\textbf{SECS$(M,I_\mathrm{max})$}}
\addlegendimage{empty legend}
\addlegendentry{}
\addlegendimage{empty legend}
\addlegendentry{}
\addlegendimage{color=anthrazit, mark=*, mark size=1pt, solid, line width=0.9pt, mark options={solid, fill=white}}
\addlegendentry{BP-10 + OSD-1}
\addlegendimage{color=darkgreen, mark=square*, mark size=1pt, solid, line width=0.9pt, mark options={solid}}
\addlegendentry{ensemble-$M$ BP-$I_\mathrm{max}$}
\addlegendimage{color=mittelblau, mark=diamond*, mark size=2pt, solid, line width=1.2pt, mark options={solid, fill=white}}
\addlegendentry{\shortstack[l]{\textbf{SECS$(M,I_\mathrm{max})$}\\+ OSD-1 [proposed]}}
\addlegendimage{empty legend}
\addlegendentry{}
\addlegendimage{empty legend}
\addlegendentry{channel:}
\addlegendimage{color=anthrazit, mark=x, mark size=2pt, dash dot, line width=0.9pt, mark options={solid}}
\addlegendentry{OSD-1}
\addlegendimage{color=anthrazit, mark=none, dotted, line width=1.0pt}
\addlegendentry{BCH: ML \cite{kldatabase}}
\addlegendimage{color=anthrazit, mark=none, dotted, line width=1.0pt}
\addlegendentry{PG: OSD-3 $\approx$ ML}
\addlegendimage{empty legend}
\addlegendentry{}
\addlegendimage{empty legend}
\addlegendentry{BKLC~\cite{GrasslCodeTables}:}
\addlegendimage{color=orange, mark=x, mark size=2pt, dash dot, line width=0.9pt, mark options={solid}}
\addlegendentry{OSD-1}
\addlegendimage{color=orange, mark=none, dotted, line width=0.9pt}
\addlegendentry{OSD-3}
\addlegendimage{color=orange, mark=square*, mark size=1pt, solid, line width=0.9pt, mark options={solid}}
\addlegendentry{ensemble-$M$ BP-10}
\addlegendimage{color=orange, mark=diamond*, mark size=2pt, solid, line width=1.2pt, mark options={solid, fill=white}}
\addlegendentry{\textbf{SECS}$(8,5)$ + OSD-1}
\end{axis}
\end{tikzpicture}
\caption{\ac{FER} over $E_b/N_0$ under \ac{AWGN} for the three studied
ensembles.  In all cases, the soft-combined \ac{LLR} followed by
\mbox{\ac{OSD}-1} (\ref{plot:SECS}) outperforms selection-based
decoding at a fraction of the iteration latency.
For the designed code, all ensembles use independent subsets of the weight-$8$ circulant pool, completed to full rank by higher-weight checks (see Sec.~\ref{ssec:dearch_designed} for details on construction). }
\label{fig:fer_all}
\vspace{-1em}
\end{figure*}
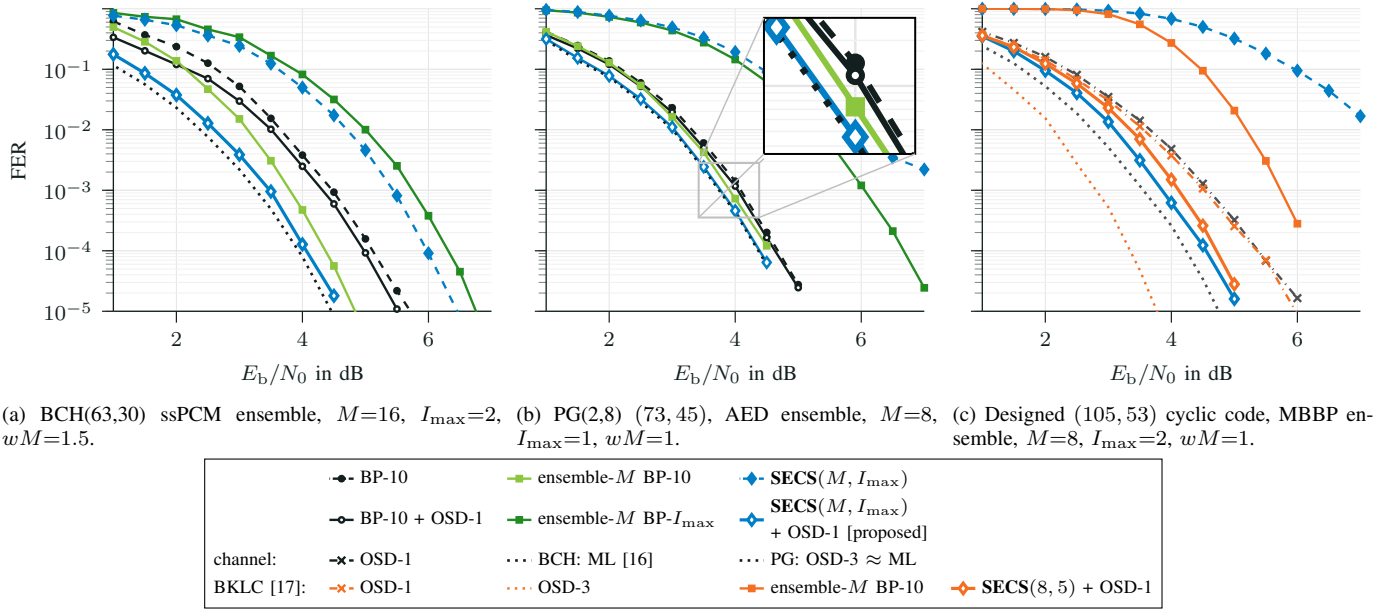
\begin{table}[!t]
\centering
\caption{Decoding budgets of the schemes in Fig.~\ref{fig:fer_all}.}
\label{tab:budget}
\scriptsize
\setlength{\tabcolsep}{3pt}
\begin{tabular}{lccl}
\toprule
scheme & \acs{BP} dec. & \acs{BP} its & post-processing \\
\midrule
\acs{BP}-10 & 1 & 10 & hard decision \\
\acs{BP}-10 + \mbox{\acs{OSD}-1} & 1 & 10 & sort, \acs{GE}, \(k{+}1\) re-enc. \\
\acs{MBBP}-\(M\) \acs{BP}-10 & \(M\) & 10 & \(M\) synd., ML-in-list \\
\acs{SECS} + \mbox{\acs{OSD}-1} & \(M\) & 1--2 & sort, \acs{GE}, \(k{+}1\) re-enc. \\
channel \mbox{\acs{OSD}-1}\,/\,-3 & 0 & 0 & sort, \acs{GE}, \(k{+}1\) / \(\mathcal{O}(k^{3})\) re-enc. \\
rebuild (Sec.~\ref{ssec:osdbybp}) & \(M\) & 22 & sort, \(2M\) synd., \acs{XOR}, ML-in-list \\
\bottomrule
\end{tabular}
\vspace{-1.5em}
\end{table}

This section evaluates the proposed \ac{SECS} on the three codes
introduced above.
All results use flooding \ac{BP}, and all schemes share the same noise realizations at the same \ac{SNR} point within one figure.
Table~\ref{tab:budget} lists the per-scheme decoding budgets.
Sec.~\ref{ssec:osdbybp} removes the \ac{GE} from the post-processing.

\subsection{BCH \((63,30)\)}
We first consider the \((63,30)\) \ac{BCH}
code~\cite{hocquenghem1959codes,bose1960class}, decoded with an
\ac{ssPCM} ensemble (see Sec.~\ref{sec:prelims}) \cite{shen2025sspcm}.
Fig.~\ref{fig:fer_bch63} compares the \ac{ssPCM} ensemble (\(M{=}16\),
tapped after \(I_\mathrm{max}{=}2\) iterations) against single-decoder
and hard-out ensemble baselines (with \ac{ML} selection).
At this low iteration count, the hard decision on the \ac{SECS} output
already outperforms the ensemble with \ac{ML} selection.
The larger gain, however, comes from the post-processing on the ordered \acp{LLR}: \mbox{\ac{OSD}-1} on
\(\Lcomb\) gains roughly two decades over \mbox{\ac{OSD}-1} on a single
\ac{BP} posterior and approaches the \ac{ML} performance of the
\ac{BCH} code.
At this rate, \ac{OSD} performance is limited by the quality of its
ordering.
The combined ordering is substantially better than both the posterior
of a converged single \ac{BP} decoder and the \ac{ML}-selected output
of a converged ensemble.

\subsection{PG(2,8)}

The \((73,45)\) projective-geometry code admits an overcomplete
\ac{PCM} of \(73\) lines.
A rank-preserving (non-overcomplete) part of it, permuted under the automorphism group,
yields eight \ac{AED} ensemble members.
Fig.~\ref{fig:fer_pg28} shows the resulting \ac{FER} performance.
We combine the eight members after a \emph{single} min-sum iteration,
followed by \mbox{\ac{OSD}-1}.
This outperforms the same members run to ten iterations with \ac{ML}
selection, at one tenth of the iteration latency.
Up to \(4\,\)dB, it also tracks the channel \mbox{\ac{OSD}-3}
reference
closely.
The strongest single-decoder baseline is \mbox{\ac{BP}-10} on the full overcomplete matrix, which, even with \mbox{\ac{OSD}-1} post-processing, cannot reach the performance of the \ac{SECS}.

\subsection{A Search-Designed $(105,53)$ Code}
\label{ssec:dearch_designed}

\begin{figure}[t]
  \centering
  \resizebox{0.95\linewidth}{!}{\input{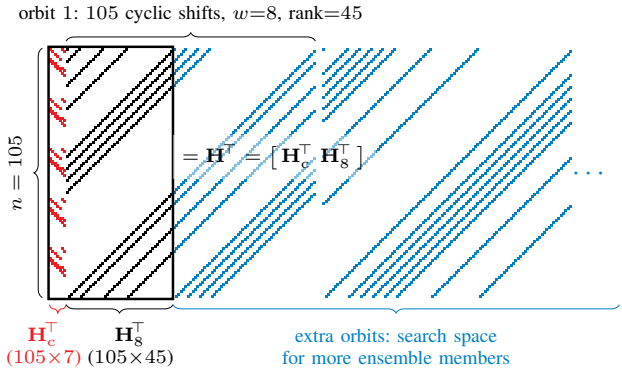}}
  \vspace{-1em}
  \caption{Rank-colored transposed overcomplete parity-check pool of the search-designed \((105,53)\) cyclic code. Black columns form a greedy independent basis; blue columns are redundant cyclic shifts.  The seven completion columns of Hamming weight $15$ (red) extend the pool to full rank $n{-}k{=}52$.}
  \label{fig:cyclic105_oc_matrix}
  \vspace{-1em}
\end{figure}

The preceding results raise the question whether codes can be
constructed specifically for this decoder.
Its main requirement is a rich supply of low-weight parity checks.
We searched cyclic codes of length \(63\)--\(105\) (rate
\(0.40\)--\(0.75\)) for \emph{abundant low-weight dual orbits}: in a
cyclic code, every dual codeword of weight \(w\) contributes its full
\(n\)-shift orbit of weight-\(w\) checks.
Some codes offer good dual orbits, but poor distance properties.
If \(x^d-1\) divides the generator polynomial, light dual orbits are
plentiful, but the code inherits low-weight codewords.
Hence, we verify the minimum distance of every candidate by
information-set decoding.

The strongest survivor is a \((105,53)\) code with \(d_{\min}{=}10\),
below the best known \(18\) for these parameters.
This is a consequence of the search objective: we optimize the supply
of low-weight dual codewords, i.e., \ac{SECS}-decodability under the
latency constraint, rather than \ac{ML} potential.
The two properties compete.

The ten weight-\(8\) dual orbits of the code form a \(1050\)-row
circulant pool of rank \(45\).
Seven dual codewords of Hamming weight \(15\) complete it to the full
rank \(n{-}k{=}52\) (Fig.~\ref{fig:cyclic105_oc_matrix}).

Each member runs a single iteration of \ac{BP}, and the outputs are combined with the mean rule (\(wM{=}1\)).
The generator polynomial has roots \(\alpha^{j}\) for all \(j\) in the
\(2\)-cyclotomic cosets of \(\{0,1,3,5,11,25,45\}\) modulo \(105\),
where \(\alpha\) is a primitive \(105\)th root of unity.

Fig.~\ref{fig:fer_c93} quantifies what the search sacrifices.
As a reference, we use the  \((105,53)\)
\ac{BKLC}~\cite{GrasslCodeTables} with \(d_{\min}{=}18\).
Under equal \mbox{\ac{OSD}-3} decoding, the \ac{BKLC} is
\(0.7\)--\(0.9\,\)dB better than our designed code at \ac{FER} below \(10^{-3}\).
This potential, however, is not accessible to iterative decoding.
An \mbox{\ac{MBBP}-8} ensemble of \ac{ssPCM} graphs, built from the
weight-\(19\) dual codewords of the \ac{BKLC}, remains above \(10^{-2}\)
until past \(5\,\)dB -- although the same construction succeeds on the
\ac{BCH} codes.
Soft combining does not rescue it either.
The deep \ac{ssPCM} members deliver no extrinsic information within the
first two iterations.
Even at five iterations, the \ac{SECS} improves over channel
\mbox{\ac{OSD}-1} only marginally.
The designed code trades this margin for a dual spectrum that a
single-iteration ensemble can exploit.
Hence, \ac{SECS} operates closer to \ac{ML} on the designed code and
outperforms the \ac{BKLC} in this complexity-limited regime.

\subsection{OSD from BP Hardware}
\label{ssec:osdbybp}

The pipeline of Fig.~\ref{fig:arch} can also be read as an \ac{OSD}
whose expensive stages are realized by \ac{BP} hardware.
A conventional \mbox{\ac{OSD}-1} implementation requires reliability
sorting, a per-frame \ac{GE} to determine the \ac{MRB}, and \(k+1\)
re-encodings.
The data-dependent \ac{GE} can impose a complexity and latency floor.
This has motivated adaptive methods that reduce or omit it whenever
possible~\cite{yue2022adaptive_ge}.
For \ac{BCH} codes, Yang et al.~\cite{yang2025bch_osd_no_ge} avoid
online \ac{GE} by exploiting the Reed--Solomon parent-code structure to
generate candidates from systematic generator matrices.
In contrast, our \ac{GE}-free variant uses the available \ac{BP}
decoder for candidate completion.
It is not limited to algebraic constructions, but does benefit from the cyclicity of the code.

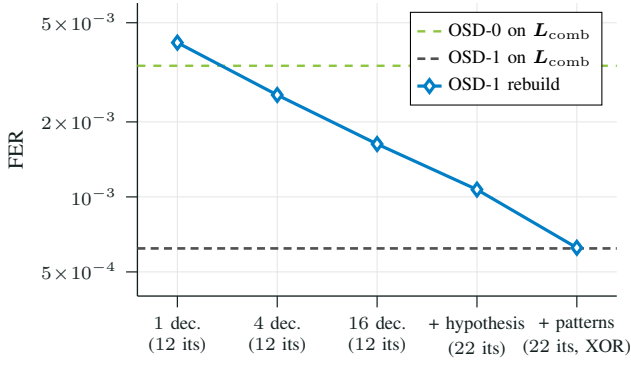
\begin{figure}[!t]
\centering
\begin{tikzpicture}
\begin{axis}[
    width=.9\linewidth,
    height=0.62\linewidth,
    ymode=log,
    log origin=infty,
    ymin=4e-4,
    ymax=6e-3,
    ytick={5e-4,1e-3,2e-3,5e-3},
    yticklabels={\(5\!\times\!10^{-4}\),\(10^{-3}\),\(2\!\times\!10^{-3}\),\(5\!\times\!10^{-3}\)},
    xmin=-0.4,
    xmax=4.4,
    xtick={0,1,2,3,4},
    xticklabels={%
      \shortstack{$1$ dec.\\($12$ its)},%
      \shortstack{$4$ dec.\\($12$ its)},%
      \shortstack{$16$ dec.\\($12$ its)},%
      \shortstack{+ hypothesis\\ ($22$ its)},%
      \shortstack{+ patterns \\($22$ its, XOR)}},
    x tick label style={font=\scriptsize, anchor=north},
    grid=both,
    major grid style={draw=hellgrau!85, line width=0.36pt},
    minor grid style={draw=hellgrau!45, line width=0.22pt},
    axis x line*=bottom,
    axis y line*=left,
    axis line style={draw=anthrazit, line width=0.50pt},
    tick align=outside,
    tick style={draw=anthrazit, line width=0.45pt},
    ticklabel style={font=\scriptsize, text=anthrazit},
    label style={font=\footnotesize, text=anthrazit},
    xlabel={},
    ylabel={FER},
    legend cell align={left},
    legend image code/.code={
        \draw[mark repeat=2, mark phase=2, #1]
            plot coordinates {(0cm,0cm) (0.16cm,0cm) (0.32cm,0cm)};
    },
    legend style={
        draw=black,
        fill=white,
        at={(0.97,0.97)},
        anchor=north east,
        legend columns=1,
        font=\scriptsize,
        inner xsep=2pt,
        inner ysep=2pt
    },
]

  \addplot+[color=apfelgruen, dashed, line width=0.9pt, mark=none]
    coordinates {(-0.4,0.0033594) (4.4,0.0033594)};
  \addlegendentry{OSD-0 on \(\Lcomb\)}

  \addplot+[color=dunkelgrau, densely dashed, line width=1.0pt, mark=none]
    coordinates {(-0.4,0.00062125) (4.4,0.00062125)};
  \addlegendentry{OSD-1 on \(\Lcomb\)}

  \addplot+[color=mittelblau, solid, line width=1.2pt, mark=diamond*,
    mark size=2.4pt, mark options={solid, fill=white}]
    coordinates {(0,0.00416) (1,0.002565) (2,0.00163) (3,0.00107) (4,0.000625)};
  \addlegendentry{\ac{OSD}-1 rebuild}

\end{axis}
\end{tikzpicture}
\vspace{-1em}
\caption{Step-by-step \ac{FER} of the \ac{GE}-free \mbox{\ac{OSD}-1}
  rebuild on \ac{BCH}(63,30) at \(3.5\)\,dB (\(M{=}16\) \ac{ssPCM} BP-2
  \ac{SECS} front end, \(wM{=}1.5\).  The first three stages scale the completion-ensemble size at
  constant latency; the hypothesis round and the pattern moves close the
  remaining gap to \mbox{\ac{OSD}-1}.}
\label{fig:readouts}
\vspace{-1.5em}
\end{figure}

The most effective follow-up stage to the \ac{SECS} that we found is a
rebuild of \mbox{\ac{OSD}-1} itself from the available \ac{BP} decoder,
needing neither the \ac{GE} nor the re-encodings.
The \(k\) most reliable positions of \(\Lcomb\) are saturated and the eight weakest positions are erased.
The \(M{=}16\) \ac{SECS} \ac{BP} decoders complete this input in two ten-iteration rounds. At first unchanged on all sixteen graphs, and then once with a weight-one sign hypothesis on each of the sixteen least reliable saturated positions.
A final \emph{pattern-move} stage adds, where it improves the metric, a precomputed low-weight codeword: \(24\) stored orbit representatives expand under the cyclic and Frobenius automorphisms to all \(8055\) codewords of weight \(\le14\); the metric change of a move 
is an exact sum over its support, and the stage triggers on fewer than \(0.1\,\%\) of frames.
Fig.~\ref{fig:readouts} shows the resulting \ac{FER} step by step on
\ac{BCH}(63,30). The performance matches \mbox{\ac{OSD}-1} on the same input at a total latency of \(22\) \ac{BP} iterations.

\section{Discussion and Conclusion}
\label{sec:conclusion}

We introduced the \ac{SECS}, a combining stage for \ac{BP} ensemble
decoders that fuses the clipped extrinsics of all members after one or
two iterations and passes the combined \ac{LLR} to a post-processing stage.
The stage exploits a simple property of unconverged \ac{BP}: shallow,
diverse members correct little, but sort exceptionally well.
On all three studied codes, the \ac{SECS} with \mbox{\ac{OSD}-1}
operates within small factors of the \ac{ML} reference and outperforms
ten-iteration selection ensembles at one fifth to one tenth of the
iteration latency.

The gain requires genuine graph diversity; its source -- automorphisms,
multiple bases, or \ac{ssPCM} constructions -- is secondary.
Optimized \ac{LDPC} codes offer little such diversity: their
only light dual codewords are the designed rows, so same-graph
ensembles built by perturbation share most failure events and gain
correspondingly little from combining.
The ordered-reliability profile and the estimator~\eqref{eq:ferest}
identify the favorable regime -- sloped per-bit \ac{BER} and no \ac{BP} overconfidence plateau.

The designed \((105,53)\) code shows that the diversity requirement is constructive.
Trading minimum distance \(18\) for \(10\) buys ten weight-\(8\) dual orbits, and the resulting code decodes within \(2\)--\(2.5\times\) of its \ac{ML} performance at a single iteration, while the best known code of equal parameters leaves its superior distance locked behind a complex decoding challenge.
Additionally, with help of an \ac{OSD}-like post-processing whose ordering and re-encoding are supplied by \ac{BP} hardware, the complete pipeline requires no \ac{GE}.
Future work includes automorphism-based member construction for
designed codes, design searches at longer block lengths, and a hardware
elaboration of the single-iteration combining datapath.

\bibliographystyle{IEEEtran}
\bibliography{references}

\end{document}